\documentclass[12pt]{iopart}

\begin{document}

\title[Machine Learning in Astronomy and Astrophysics ]{Foreword to the Focus Issue on Machine Learning in Astronomy and Astrophysics}

\author{Giuseppe Longo$^{1}$, Erzs\'ebet Mer\'enyi$^{2}$, and Peter Ti\v{n}o$^{3}$}

\address{1- Department of Physics, University Federico II in Napoli, via cintia, Napoli I-80126}
\ead{longo@na.infn.it}
\address{2- Department of Statistics, Rice University, Houston, Texas, USA}
\ead{erzsebet@rice.edu}
\address{3- School of Computer Science, The University of Birmingham, Birmingham, UK }
\ead{P.Tino@cs.bham.ac.uk}

\vspace{10pt}
\begin{indented}
\item[May 2019]
\end{indented}

%
%
%
%
%

\section{The need for machine learning in astronomy}

Astronomical observations already produce vast amounts of data through a new generation of telescopes (Atacama Large Millimeter Array (ALMA), Jansky VLA) and through large surveys (e.g., SDSS \cite{york2000}, ZTF \cite{bellm2014}, PanSTARRS \cite{kaiser2010}, VLT Survey Telescope - VST, and many others) that cannot be analyzed manually. Next-generation telescopes such as the Large Synoptic Survey Telescope (LSST \cite{ivezic2008}) and the Square Kilometer Array (SKA \cite{dewdney2009}) are planned to become operational in this decade and the next, and will increase the data volume by many orders of magnitude. 
The increased spatial, temporal and spectral resolution afford a powerful magnifying lens on the physical processes that underlie the data but, at the same time, generate unprecedented complexity hard to exploit for knowledge extraction. It is therefore imperative to develop machine intelligence, machine learning (ML) in particular, suitable for processing the amount and variety of astronomical data that will be collected, and capable of answering scientific questions based on the data \cite{baron2019}.

Astronomical data exhibit the usual challenges associated with “big data” such as immense volumes, high dimensionality, missing or highly distorted observations. In addition, astronomical data can exhibit large continuous observational gaps, very low signal-to-noise ratio and the need to distinguish between true missing ({\it i.e.}, non-collected) data and non-detections ({\it i.e.}, due to upper limits). There are strict laws of physics behind the data production which can be assimilated into ML mechanisms to improve over general off-the-shelf state-of-the-art methods. An additional peculiarity is that these large and heterogeneous data sets \cite{pesenson2010} need to be simultaneously queued, merged and mined by many independent groups of researchers, posing problems not common in many other application domains. In this context it is important to mention the crucial role played by the International Virtual Observatory Alliance - IVOA \cite{IVOA} which aims at a seamless access to astronomical data and focuses on the standardization of data and metadata, data exchange methods, and to the implementation of a registry, which lists available services and what can be done with them. 

Significant progress in the face of these challenges can be achieved only via the new discipline of Astroinformatics \cite{IAU325}: a symbiosis of diverse disciplines, such as ML, probabilistic modelling, astronomy and astrophysics, statistics, distributed computing and natural computation. The importance of the task resulted in the emergence of (often large) interdisciplinary collaborations. 
This Focus Issue offers a sample of progress from recent years in enabling scientific discoveries using ML, by 69 authors representing 15 countries, from 6 continents.  

Machine intelligence provides capabilities for discovering intricate relations and extracting information, making accurate inference from complex multi-dimensional data; and also the capability to process large amounts of data fast. This collection of papers focuses primarily on the former capability, although speed considerations also surface in most works.  A noteworthy point is that the majority of the ML methods in these papers are Artificial Neural Networks (ANNs), including Multi-Layer Perceptrons with Back Propagation (MLP / BP), Self-Organizing Maps (SOMs) and Deep Learning networks (Convolutional and Recurrent Convolutional Neural Nets). This confirms a trend already noted by many: the slow adoption by the astronomical community of less well-known but possibly more effective ML methods than those that have traditionally been used. In the recent past, most works focused only on supervised classification/regression methods applied to a handful of problems, mainly due to the lack of suitably large bases of knowledge ({\it i.e.}, template data sets) on which to train the models. The broad spectrum of applications presented in this special issue shows instead that ML methods are becoming widely used and that new applications arise every time a new data set is made publicly available. In what follows we briefly summarize the papers contained in this Focus Issue, grouping them by the astronomical problems targeted, with the aim of emphasizing the spectrum of ML applications.

\section{Astronomical applications of ML in this issue}

Classification of astronomical sources is one of the central problems addressed in this issue. The respective approaches can be divided into two broad groups. One involves ML methods that aim to reproduce, on large data sets of objects, predefined classification schemes specified on subjective grounds by human experts (supervised methods). Methods in the other group aim at obtaining an unbiased classification based on the statistical properties of the objects (unsupervised clustering methods). Both approaches have their pro's and con's and are useful in their own terms. 

\begin{itemize}
\item In \textit{Deep Learning for Image Sequence Classification of Astronomical Events} Carrasco-Davis {\it et al.}   propose a novel approach to the classification of variable stars and transients (and in general, of astronomical objects in large data streams). They bypass the traditional pre-processing of temporal sequences of images for extracting features --- often light curves, further reduced to parameters derived from the light curves --- that allows to classify much lower-dimensional signals than the image sequence. Instead, they train a Recurrent Convolutional Neural Network (RCNN) to extract the relevant features (latent variables) directly from the raw data, then use the latent feature vector as input to the classifier component of the network. This allows automatic, data-driven feature extraction without the need of computing difference images, making model assumptions, and corrections throughout the various steps; and helps avoid errors that may be introduced by computing light curves. The authors also describe a method to simulate synthetic image sequences based on the instrument and observation characteristics of a given experiment (the HiTS survey in their example), for the purpose of training the RCNN (which requires a large number of labeled samples). The RCNN can subsequently be fine-tuned to the distribution of data actually measured from the assumed experiment, by using a small number of the real labeled samples. 
\end{itemize}

\noindent A more traditional approach to classification problems, based on morphological features is the topic of several papers. 

\begin{itemize}
\item \textit{Multiband Galaxy Morphologies for CLASH: a Convolutional Neural Network Transferred from CANDELS} by Perez-Carrasco {\it et al.} uses a convolutional neural network (CNN) to classify galaxies into broad morphological classes - spheroid, disk, irregular, point source, or “unclassifiable”.  The authors use a large collection of images of over 8400 galaxies from the Cluster Lensing And Supernova survey with Hubble (CLASH) in 16 photometric bands, from ultra-violet to near-infrared. A big challenge in such studies is the knowledge of the ``truth'' labels attached to the galaxy images and used in the processes of training and evaluating the classifiers, because the morphological labeling provided by humans may not be perfect. The authors take great care to address this issue. In particular, five experts were used to provide labels for 100 randomly selected galaxies in each of the 16 filters. This subset of galaxies with highly reliable labels is then used to evaluate the classifier performance in greater detail and fine tune the deep network. 

\item Self-Organizing Maps (SOMs) play the key role in \textit{Radio Galaxy Zoo: Knowledge Transfer  Using Rotationally Invariant Self-Organizing Maps} by Galvin {\it et al.} to facilitate automatic labeling (classification) of a large number of galaxies according to their morphological complexity. A topological feature map of Radio Galaxy Zoo (RGZ, \cite{banfield2015}) images is learned, and the SOM prototypes (weight vectors) are subsequently labeled according to the plurality of the labeled samples in their receptive fields. The labeled training samples are objects that have high consensus among citizen scientists who labeled the RGZ images. The remaining citizen-scientist-labeled images are then mapped to the learned SOM prototypes to see how they align with the topology learned from the high-consensus labeled images; as well as to classify them according to their best-matching SOM prototypes. For the labeling / classification of RGZ objects they use SOM heat maps as input to a Quantile Regression Forest (QRF), where a heat map is the matrix of similarities between an input image and each of the SOM prototypes.  The QRF then performs a probabilistic assignment into classes. One noteworthy aspect is that for the winner selection in SOM learning a modified Euclidean distance is utilized, in the framework of the PINK package, which incorporates affine transformations. This accounts for rotated versions of source images, providing rotational invariance in a more elegant way than the customary generation and storage of transformed images.

\item Ralph {\it et al.} in \textit{Radio Galaxy Zoo: Unsupervised Clustering of Convolutionally Auto-encoded Radio-astronomical Images} also use SOMs to cluster radio astronomy images from the RGZ. The paper is part of the ongoing preparation for the future SKA radio surveys and is focused on demonstrating that a very much needed increase in computational efficiency (speed) can be achieved via intelligent compression of the data fed to the SOM. 
In their work they use compressed versions of the raw images (vectors of extracted latent variables) obtained by a (supervised) CNN autoencoder, which proves capable of preserving the relevant information for accurate representation of the manifold structure by the SOM. A very interesting aspect of this paper is that SOM knowledge is also viewed as a way to transform a discrete (hard) classification task into a continuous regression problem, which could open the way to a more physical classification of the radio morphologies. This aspect is subject of future works by the same group. 

\item In a similar line of work, \textit{Unsupervised Classification of Galaxies. I. ICA Feature Selection}, Chattopadhyay {\it et al.} tackle the problem of categorizing galaxies using an unsupervised approach --- K-means clustering --- for over 300,000 galaxies from the Value Added Galaxy Catalogue (VAGC), and utilizing ICA feature selection prior to clustering to reduce the dimensionality of the data. Importantly, the authors use 49 measurable physical attributes including photometry, spectroscopy, morphology, chemical composition and kinematics, but no non-observable features. Their results are very encouraging since they succeed in identifying clusters of objects with different Sersic index, active or non-active, and with different star formation history, etc., thus paving the way toward a physically based classification scheme. We wish to note also that unsupervised approaches do not suffer from the label biases which are always present in attempts to reproduce galaxy classification using supervised methods and therefore are more likely to be able to capture --- discover --- physical phenomena (such as the presence of an AGN) which are often missed by traditional morphology.
\end{itemize}

\noindent Often, in huge collections of data, one would like to discover unexpected phenomena that stand out as not aligned with the general trends in the data. This is also known as ``novelty" or ``anomaly" detection. 
\begin{itemize}
\item In \textit{Identifying Complex Sources in Large Astronomical Data Using a Coarse-grained Complexity Measure}, Segal {\it et al.} explore the possibility of finding such phenomena using a detector based on a complexity measure applied to data items. In this context, data items with unusual complexity values will stand out. In particular, the authors introduce a measure on image data, termed “apparent complexity”, that is based on a combination of information theoretic considerations (entropy) and a smoothing function to filter out random elements in the images. The measure can be used to segment and identify “interesting” observations in huge data repositories by quantifying their morphological complexity. The method is demonstrated on the Australia Telescope Large Area Survey (ATLAS) to distinguish between images of galaxies with simple and complex morphologies. 
\end{itemize}

\noindent The evaluation of photometric redshifts is among the most common applications of ML methods in astrophysics. The number of applications is expected to drastically increase in the coming years, when new photometric and spectroscopic surveys will produce accurate and large datasets of precisely measured parameters to be used as templates. At the moment, however, photometric redshifts have become a sort of benchmark for comparing different methods and procedures. Two papers address the problem of obtaining reliable redshift estimates for the radio selected samples which will be the main target of future surveys such as EMU (Evolutionary Map of the Universe) to be performed with SKA. These papers use regression --- a supervised technique where the target variables are continuous --- such as in the case of photometric redshifts (\cite{tag}, \cite{salvato} and references therein), and of stellar formation rates \cite{stensbo}, \cite{michele} in galaxies. This is in contrast to discrete target variables, often called labels, in supervised classification. Both works address a problem commonly affecting all supervised methods, but so far little-explored in this context: how performances are affected by biases in the construction of the training sets.

\begin{itemize}
\item In \textit{A Comparison of Photometric Redshift Techniques for Large Radio Surveys}, Norris {\it et al.} study ML techniques --- k-Nearest Neighbors (kNN), Random Forests (RF) and a Multi-Layer Perceptron with Quasi Newton Algorithm for optimization (MLPQNA) --- for photometric redshift estimation for radio sources, in comparison with a classic template-fitting approach, Le Phare \cite{lephare}. The study targets the scenario near-future surveys present with tens of millions of radio sources, for which redshift determination is essential but the general lack of spectroscopic redshift measurements (only available for a fraction of the sources) necessitates estimation of the redshift from photometric measurements (available for all sources). The comparative analyses are performed in a multi-dimensional parameter space that includes low and high ranges of redshift values; quality of photometric data; photometry bands; depth of photometric survey; completeness of training data; stratification by type of source; and more, providing advice about the relative benefits of the techniques under various circumstances. 

\item In \textit{Preliminary Results of Using k-Nearest Neighbours Regression to Estimate the Redshift of Radio Selected Datasets}, Luken {\it et al.} use k-Nearest Neighbor regression to explore a different aspect of the problem. They start from a dataset obtained from the fusion of different surveys covering a large wavelength range, from the Far Infrared to the optical, and degrade it in order to match the expected shallowness of EMU. Their results confirm the need for a training set providing a good coverage of the parameter space, and provide a strong evidence that there is no need for the training and test set to be extracted from the same region of the sky, thus opening the way to the possibility of using deep, pencil beam surveys to complement spectroscopic data obtained from shallow wide field surveys.  
\end{itemize}

\noindent Machine learning can be a useful tool in the model-based exploration of astronomical data. 

\begin{itemize}
\item 
A nice example of this scenario is the paper \textit{Deep Learning Applied to the Asteroseismic Modeling of Stars with Coherent Oscillation Modes} by Hendriks and Aerts. The authors use a deep learning network to interpolate between nodes of a huge grid of stellar evolution models. The grid of models is realized as a grid in the parameter space that can be high-dimensional, prohibiting any fine grid-based exploration of the model space. The deep network generalizes to models between the grid points and thus can help to adequately parametrize forward asteroseismic modelling. Otherwise, we would be restricted to models corresponding to the (possibly quite sparse) set of grid points. The deep network is trained on a grid of astero-seismological models (parameters) of intermediate- and high-mass stars to predict the frequencies of their coherent oscillation modes. The authors employ a genetic algorithm to identify appropriate regions for the stellar parameters based on the oscillation mode frequencies. 
\end{itemize}

\noindent An entirely different topic, Adaptive Optics (AO) is the subject of one paper.
\begin{itemize}
\item 
\textit{Experience with Artificial Neural Networks Applied in Multi-Object Adaptive Optics} by G\'omez {\it et al.} addresses AO with Artificial Neural Networks, specifically Multi-Layer Perceptrons (MLPs) with Back Propagation (BP). The paper reviews, in the context of Multi-Object Adaptive Optics, the state-of-the-art in Artificial Intelligence dominated by MLP applications, compared to more traditional methods for fast reconstruction of the wavefront of light after having been deformed by Earth' atmosphere. Simulated atmosphere and “on-sky” (real data) scenarios are discussed using the CANARY flexible AO demonstration bench and the CARMEN (Complex Atmospheric Reconstructor based on Machine LEarNing) framework involving MLP/BP ANNs. Relative merits of different ANN models for different atmospheric conditions are presented.  Extensions of CARMEN to scales such as those of the European Extremely Large Telescope and the Thirty Meter Telescope are then elaborated including computation (recall) speed, aiming at near-real time reconstruction; concluding with a next generation of CARMEN that further improves the reconstruction by using a CNN. 
\end{itemize}

\noindent More general ML issues are targeted by Buchner, and by Vilalta {\it et al.}, namely, sampling and transfer learning. In the age of big data, model-based ML performed in the Bayesian framework is often pushed aside for pragmatic reasons. First, while the Bayesian framework is principled and allows for proper treatment of uncertainties (including, {\it e.g.}, different types of noise models), it unfortunately is “computationally hungry”. Dealing with huge data sets and many models can quickly become infeasible. Second, it is often the case that point estimates of semi-parametric ML models, such as deep networks, perform empirically very well when trained on adequately large data sets and are increasingly available in ready-to-use packages. This is fine as long as one's goal is classification of astrophysical objects of interest, or predictions on such objects. This is, for example, the case in the 
paper by Perez-Carrasco {\it et al.} (this issue), where the authors use a convolutional deep network to classify galaxies into a set of predefined morphological classes. However, we would often like to include, in learning from the data, prior knowledge which is naturally expressed as interpretable physical models. In that case the use of deep networks can be problematic. 
\begin{itemize}
    \item 
The paper \textit{Collaborative Nested Sampling: Big Data vs. Complex Physical Models}  by Buchner proposes an innovative and efficient method for sampling in the model space (and thus enabling Bayesian framework) in the case of very large datasets. The main principle of this approach is that (especially in the initial stages) the sampling regions can be quite similar across similar data sets and so rejection sampling from the union of such contours can be performed. 

\item
Transfer learning is another topic of growing importance in ML, useful when expensive learning of a complex model is performed using a data set from one domain and then the model is “exported” to another. In Vilalta {\it et al.} \textit{A General Approach to Domain Adaptation with Applications in Astronomy}  the authors use as astronomical template case the identification of Type Ia Supernovae. This represents a problem of paramount importance for the emerging field of Time Domain Astronomy and especially for the scientific exploitation of LSST data which is expected to produce $10^{6-7}$ transients per night, with only a small fraction of Type Ia supernovae among them.  In the paper the authors show how to transfer learning from spectroscopic data to photometric data. The paper is innovative from the algorithmic point of view (a new approach to domain adaptation which does not depend on the proximity to the source nor on the target distribution) as well as for the astronomical implications: the possibility to increase performance and reduce computational costs in tackling complex problems in very large data sets.  
\end{itemize}

\section{Perspectives}

In less than a decade, the application of ML methods to astronomical problems has become a widespread practice. This growth has been triggered mainly by the huge, comprehensive data sets provided by modern multi-band, multi-epoch digital sky surveys, as well as by the standardization of heterogeneous data induced by the creation of the Virtual Observatory. It is also clear that the most exciting innovations come from large interdisciplinary groups where the expertise of data scientists and domain experts are combined. In the near future, this kind of virtuous collaboration between experts from different disciplines will probably allow to overcome some of the main problems the astronomical community is currently facing. 

One of these problems is that most supervised learning algorithms are not fine-tuned to deal with astronomical datasets, since most assume that all measured features are of the same quality, and that the labels are correctly reflecting the truth \cite{baron2019}.  However, astronomical data are sparse andhttps://v2.overleaf.com/project/5cdce16a4a8bb2207ced3d9d noisy, and in most cases, the targets in the training sets are derived by human experts, thus prone to inconsistencies. This implies that in most applications, supervised learning algorithms perform well when applied to high signal-to-noise ratio datasets, or to (the few) datasets characterised by uniform noise properties. It is therefore imperative to modify existing tools and/or to devise new algorithms, capable of taking into account uncertainties during the model construction and to predict uncertainties based on the intrinsic properties of the objects in the sample and on their measurement uncertainties \cite{metafor}. While external domain knowledge and scientific judgment can be infused to weight the reliability of labels (as in the case of citizen-labels of the RGZ); classifiers whetted in other, closely related  big-data domains such as terrestrial and planetary hyperspectral imaging can be helpful for coping with data-dependent noise, labeling errors, and severely biased training sets \cite{Merenyietal2014}. 
Principled construction of classifiers that can handle label noise is increasingly studied in ongoing ML work, {\it e.g.,} \cite{Bootkrajang14}, including incorporation of knowledge about the nature of the process that corrupts the labels, for enhancing the robustness of the classifier estimation. 

Another issue that will prove crucial in future applications is knowledge transfer. Models trained on a specific data set usually fail to generalize when applied to another dataset with even slightly different properties. For the exploitation of future surveys it will be un-thinkable to train algorithms on the final data releases, therefore algorithms will need to be trained while the data are still being acquired.  Transfer learning, which is widely discussed in the ML literature but still seldom applied in Astronomy, will facilitate this.

A third aspect which needs to be considered is that the performance of all supervised learning algorithms strongly depends on the number of features that are selected, since most are either not designed to work with hundreds or thousands of features (will break down mathematically), or an excessive number of features with redundant information decreases the discrimination capability of the algorithm.  Traditionally, dimensionality reduction was achieved relying on the choices by experts, but this approach is now being superseded by automatic techniques \cite{disanto2018,michele}, and by exhaustive, albeit computationally intensive, approaches \cite{pol1}. Alternaively, ML can also be applied to derive, for a given classification goal in a many-class scenario in high-dimensional feature spaces, the relative importance of the measured features  \cite{MendenhallMerenyi2008}. In contrast to methods that return reduced (latent) features in transformation spaces ({\it e.g.}, PCA or deep convolutional networks), this provides (non-linear) dimensionality reduction in the original feature space, offering a readily interpretable subset of the original physical features. More, computationally efficient algorithms for feature selection are badly needed to cope with the huge data volumes and data streams which are foreseen for the near future. 

An extremely important resource for scentific inquiry in astronomy and astrophysics, not represented in this issue, is  spectral image data with hundreds or thousands of image planes. These ``deep data" are increasingly available from advanced telescope systems, ALMA currently representing the frontier. While ``wide" sky surveys such as LSST cover large segments of the sky with high cadence in a few broad optical bands, ALMA and VLA data provide immensely rich compositional, kinematic, and other information for targeted objects such as protoplanetary disks or molecular clouds, offering high potential for discovery. ALMA and VLA also image objects in multiple molecular lines which can be combined (stacked) for further discovery power. However, current standard analysis tools fall short of fully exploiting these data, even from a single molecular line. The large number of image bands (large feature dimension with substantial frequency gaps), complex, irregular cluster structure inherent in such data exceed the capabilities of many unsupervised (clustering, anomaly detection) ML methods. Recent collaborative work that successfully extracts physical phenomena using all channels from multiple stacked ALMA lines of a protoplanetary disk \cite{Merenyietal2016} is spearheading developments in this direction. 

Finally, we want to underline that interpretability is a crucial requirement in many ML applications. One possibility of achieving it is through a creative blending of model based approaches with data driven automated learning. For example, one can impose a latent flux from the common source (quazar)  as an integral part of the overall machine learning model for resolving the delay in gravitationally lensed images \cite{Cuevas-Tello10,Otaibi16}. Another example is the imposition of a smooth latent low-dimensional structure in the space of interpretable models (as opposed to formulating it in the data space) to discover the pivotal degrees of freedom with physical meaning buried in the observed data \cite{Gianniotis08,Gianniotis09}. It is important to strive to construct learning methods that, while learning in a data driven manner, are able to provide scientifically interpretable and physically meaningful predictive outputs. A very active line of this research aims at making  deep neural networks interpretable, see, {\it e.g.}, \cite{MONTAVON20181}.\\

This Focus Issue –-- as far as we know --- is a first of its kind to appear in an astronomical journal, which posed special challenges. The authors are to be commended for a great effort to make their work accessible to both the astronomy and the machine learning communities while striving to follow the standards of both. We hope that this relatively small collection will encourage similar interdisciplinary issues soon.

\vspace{0.5in}
\hfil\hfil Giuseppe Longo, Erzs\'ebet Mer\'enyi, Peter Ti\v{n}o

\hspace{3.0in} Guest Editors\\

\section*{Acknowledgments}
GL wishes to thank G. Djorgovski, D. Baron, M. Brescia for many stimulating discussions and useful suggestions. This work was partly sponsored by the SUNDIAL Marie Curie ITN of the Horizon 2020 program. \\

\end{document}